# Band Engineering of Dirac Semimetals using Charge Density Waves

*Shiming Lei Samuel M. L. Teicher Andreas Topp Kehan Cai Jingjing Lin Guangming Cheng Tyger H. Salters Fanny Rodolakis Jessica L. McChesney Saul Lapidus Nan Yao Maxim Krivenkov Dmitry Marchenko Andrei Varykhalov Christian R. Ast Roberto Car Jennifer Cano Maia G. Vergniory N. Phuan Ong Leslie M. Schoop*

S. Lei, K. Cai, T.H. Salters, R. Car, L.M. Schoop
Department of Chemistry, Princeton University, Princeton, New Jersey 08544, USA
Email Address: lschoop@princeton.edu
S. M. L. Teicher
Materials Department and Materials Research Laboratory, University of California, Santa Barbara, California 93106, US
A. Topp, C. Ast
Max-Planck-Institut für Festkörperforschung, Stuttgart, 70569, Germany
J. Lin, R. Car, N.P. Ong
Department of Physics, Princeton University, Princeton, NJ 08544, USA
G. Chen, N. Yao
Princeton Institute for Science and Technology of Materials, Princeton, New Jersey 08544, United States
F. Rodolakis, J.L. McChesney, S. Lapidus
Argonne National Laboratory, 9700 South Cass Avenue, Argonne, IL 60439, USA
M. Krivenkov, D. Marchenko, A. Varykhalov
Helmholtz-Zentrum Berlin für Materialien und Energie, Elektronenspeicherring BESSY II, Albert-Einstein-Straße 15, 12489 Berlin, Germany
J. Cano
Department of Physics and Astronomy, Stony Brook University, Stony Brook, New York 11974, USA
Center for Computational Quantum Physics, The Flatiron Institute, New York, New York 10010, USA
M. Vergniory
Donostia International Physics Center, 20018 Donostia-San Sebastian, Spain and IKERBASQUE, Basque Foundation for Science, Maria Diaz de Haro 3, 48013 Bilbao, Spain

New developments in the field of topological matter are often driven by materials discovery, including novel topological insulators, Dirac semimetals and Weyl semimetals. In the last few years, large efforts have been performed to classify all known inorganic materials with respect to their topology. Unfortunately, a large number of topological materials suffer from non-ideal band structures. For example, topological bands are frequently convoluted with trivial ones, and band structure features of interest can appear far below the Fermi level. This leaves just a handful of materials that are intensively studied. Finding strategies to design new topological materials is a solution. Here we introduce a new mechanism that is based on charge density waves and non-symmorphic symmetry to design an idealized Dirac semimetal. We then show experimentally that the antiferromagnetic compound GdSb$_{0.46}$Te$_{1.48}$ is a *nearly* ideal Dirac semimetal based on the proposed mechanism, meaning that most interfering bands at the Fermi level are suppressed. Its highly unusual transport behavior points to a thus far unknown regime, in which Dirac carriers with Fermi energy very close to the node seem to gradually localize in the presence of lattice and magnetic disorder.

# 1 Introduction

Topological matter has fascinated the field of condensed matter physics in the last decade due to its potential for understanding properties of matter in a new way. Phenomena that have been discovered in topological materials include the quantum spin Hall effect [1, 2], quantum anomalous Hall effect [3], absence of back-scattering [4], ultrahigh carrier mobility and giant magnetoresistance [5, 6, 7], and topological Fermi arc states [8, 9, 10, 11]. While many topological materials, including Dirac semimetals (DSMs), have been discovered, there are only a handful that have been studied intensively [12, 13, 14]. Recently, various algorithms have been developed to scan through a large number of known non-magnetic materials, which are then catalogued in databases with respect to their topological classification [15, 16, 17]. While these efforts are a great help for gaining a general understanding about what kinds of materials are topological, coincidentally the "best" topological materials had already been discovered before. For example Sn-doped $(Bi,Sb)_2Te_2S$ is still the topological insulator with the largest band gap [18] and graphene is still the DSM with the largest range of linear band dispersion [19]. Common shortcomings of most topological materials are that their topologically relevant states are often below/above the Fermi level or that trivial states interfere with the relevant bands at the Fermi level. Furthermore, the materials search lags behind when it comes to magnetic materials, where only a few studies with limited examples exist [20, 21].

If viewed from a chemical perspective, topological band structures can be linked to delocalized chemical bonds, which often appear in compounds that are prone to undergo Peierls distortions [22, 23]. For example, the Dirac cone in graphene is chemically stabilized by its delocalized, conjugated π-electrons. Although graphene features a half-filled band, which would indicate a propensity to a charge density wave (CDW) distortion, it keeps its hexagonal symmetry. Such behavior can be understood by considering graphene's Fermi surface (FS), which only consists of isolated K and $K^t$ points and thus disfavors the FS nesting condition of CDW formation. A different class of topological semimetals (TSMs) that features delocalized bonds are square-net materials. An example is the nodal-line semimetal ZrSiS, which features a dense Si square net with half-filled $p_x$- and $p_y$-bands [24, 25]. The band structure features a diamond-shaped nodal line at the Fermi level, which will gap to a weak topological insulating state with high spin-orbit coupling (SOC), as well as four-fold degenerate Dirac nodes at the Brillouin zone (BZ) boundaries, which are robust against SOC (Figure 1). The latter are a consequence of non-symmorphic symmetry, which was previously suggested as a mechanism to design DSMs in square nets by Young and Kane [26]. For half-filled $p_x$- and $p_y$-bands, the non-symmorphically-enforced Dirac crossing can appear at or near the Fermi level, depending on the strength of next-nearest-neighbor interactions in the square net [26, 27]. A material with a "clean" (i.e. with no other interfering bands) non-symmorphically protected Dirac node at the Fermi level has not yet been achieved, to the best of our knowledge.

Square-net materials with the *MXZ* formula in space group *P4/nmm* (isostructural to ZrSiS), are known to have large chemical flexibility and importantly, the *M* site is able to incorporate rare earth elements, thus providing an opportunity to study the effect of magnetism and potentially correlations on the topological band structure. In this context, *Ln*SbTe materials (*Ln* = lanthanide) that are isostructural and isoelectronic to ZrSiS have been suggested as promising candidates as magnetic TSMs [28, 29, 30, 31, 32, 33]. However, the band structure of these materials is not as "clean" as in ZrSiS: the FS contains trivial pockets, in addition to the nodal-line states.

Recently, CDWs have gained increased interest in the context of TSMs due to their distinct role in the formation of novel topological phases [34, 35, 36, 37, 38]. It is well established that CDWs can be induced by chemical substitution in *Ln*SbTe systems [39, 32]. Here, we show how these CDWs, in combination with non-symmorphic symmetry, can be utilized to design "clean" non-symmorphic DSMs: The CDW gaps out states within the BZ, while the non-symmorphic symmetry-enforced band crossings at the BZ boundary, which is at the Fermi level in $LnSb_xTe_{2-x}$ for certain values of *x*, is unaffected. Thus, we introduce a new mechanism to design non-symmorphic DSMs, that relies on CDWs. Experimentally, we focus on $GdSb_{0.46}Te_{1.48}$ and show with angle-resolved photoemission spectroscopy (ARPES) that a non-symmorphically protected Dirac crossing appears at the Fermi level, with minimal interference from trivial bands. In addition, we reveal that $GdSb_{0.46}Te_{1.48}$ exhibits a very complex magnetic phase diagram and highly unusual transport properties.

## 1.1 Results and Discussion

Square-net materials, such as ZrSiS and GdSbTe, feature side-centered square nets as shown in Figure 1a. Such nets are commonly referred to as $4^4$-nets in crystallography literature [40, 41]. The atoms occupying the $4^4$-net (Si and Sb respectively) have six electrons, resulting in half-filled $p_x$- and $p_y$-orbitals. A tight-binding (TB) model of a two-dimensional $4^4$ net of $p_x$- and $p_y$-orbitals, similar to that in refs. [27, 23], is thus considered. The resulting TB band structure is shown in Figure 1b. Both the nodal line that defines the diamond-shaped FS (referred to as the *mirror-symmetry-protected* Dirac node (m-DN)), and the non-symmorphically enforced degeneracies at X and M (referred to as *non-symmorphically-protected* Dirac node (ns-DN)) are revealed in the model. When electrons are added to the system, the FS will become nested (Figure 1c). Therefore CDWs are expected to appear when adding electrons to the system. In accordance, CDWs have been reported in LaSe$_2$ [42], rare-earth ditellurides [39], and rare-earth tritellurides [43]. Note that the CDW appears above room temperature in these systems. For a band filling of $E = E_2$ (Figure 1b), the Fermi level crosses the ns-DN at X. In addition, the Fermi level crosses the m-Dirac nodal-line bands along Γ-X and Γ-M, but above the m-DN. With a CDW, these additional band crossings can be gapped at the Fermi level, ideally to just leave the ns-DN. To test this hypothesis, we extended the TB model to a superstructure reflecting the CDW.

Previously, we reported on the structural evolution of GdSb$_x$Te$_{2-x-\delta}$ ($\delta$ describes the vacancy concentration), with varying Sb composition $x$ [32]. The CDW wave vector $q_{CDW}$ in GdSb$_{0.46}$Te$_{1.48}$ was determined to be 0.20 r.l.u. (reciprocal lattice unit) and single crystal x-ray diffraction revealed that GdSb$_{0.46}$Te$_{1.48}$ adopts a five-fold superstructure in the orthorhombic space group *Pmmn*, in which the $4^4$-net forms zig-zag chains (Figure 1d), retaining the non-symmorphic symmetry [32]. The square net in the TB model was modeled with anisotropic nearest-neighbor-hopping parameters as well as by a five-fold increase of the unit cell in one direction (Figure 1e, more details are given in Experimental Section). To compare the resulting band structure with that of the subcell, the supercell band structure is unfolded into the subcell BZ (Figure 1f, also see Figure S1, Supporting Information). The Fermi level was set to $E = E_2$. Along Γ-X, there are now several ns-DNs that result from band folding, which appear with reduced spectral weight relative to the original one. The Fermi level cuts through the center of each of these ns-DNs. Along S-Γ, the bands that contribute to the m-DN also cross the Fermi level, because the band gaps created by the CDW appear slightly above $E_F$. However, depending on the intensity of the CDW modulation (which we have modeled by the strength of the differed hopping in the TB model), the Fermi level can reside within the gaps along S-Γ, as we will experimentally show below. Note that the hybridization between the CDW folded bands and the original bands is naturally included in the supercell tight-binding model calculation. The hybridization effect and the resulting gaps appear mostly at the supercell BZ boundary that lies within the subcell BZ. Our TB model thus shows that a CDW will generally open gaps in the band structure, but the non-symmorphically-protected Dirac cones will be preserved, as long as the CDW preserves the non-symmorphic symmetry.

We now consider the real material system GdSb$_x$Te$_{2-x-\delta}$. An illustration of the crystal structure of undistorted, tetragonal GdSbTe is shown in Figure 1g and the corresponding DFT-calculated band structure is shown in Figure 1i. Note that GdSbTe is modeled with the same orthorhombic lattice parameters as the subcell of GdSb$_{0.46}$Te$_{1.48}$ and thereby the same *Pmmn* space group, to ease comparison. The BZ for a primitive orthorhombic cell is shown in Figure 1h. In the paramagnetic state, bands are guaranteed to be four-fold degenerate (counting spin) at the X, U, Y, T, S, and R points, respectively (for more details see Supporting Information).

Figure 1j shows an illustration of the ns-Dirac nodal line along X-U, which reveals a $k_z$ dispersion. The ns-DN at X is 0.58 eV below that at U. Consequently, a Dirac crossing would appear at the FS as long as the Fermi level of GdSb$_x$Te$_{2-x-\delta}$ resides in between the energy span of X and U, which is indicated by the colored window in Figure 1i. When the Fermi level resides in the blue-shaded region, it cuts through additional bands along Γ-X-S-Γ (marked by the arrows), resulting in a hole-pocket centered at Γ. Since an inclusion of a three-dimensional hole-pocket (see Figure S2a, Supporting Information) would impede CDW formation, the doping level $x$ in GdSb$_x$Te$_{2-x-\delta}$ should be high enough to move the Fermi level to the green-shaded region, where this hole-pocket disappears.

In this work, we choose GdSb$_{0.46}$Te$_{1.48}$ as the material of focus, where the Fermi level lies in the green region and a CDW, which preserves the non-symmorphic symmetry, has been reported to exist above room temperature [32].

We performed ARPES measurements on GdSb$_{0.46}$Te$_{1.48}$ single crystals to verify the proposed mechanism. A polarized optical image of one such crystal is shown in Figure 2b. Figure 2a shows a selection of constant energy cuts from $E_F - 0.5$ eV to $E_F$, measured with a photon energy of $h\nu = 70$ eV. We deduced this photon energy to correspond to the $k_z \approx \pi/c$ plane (for photon energy dependent data see Figure

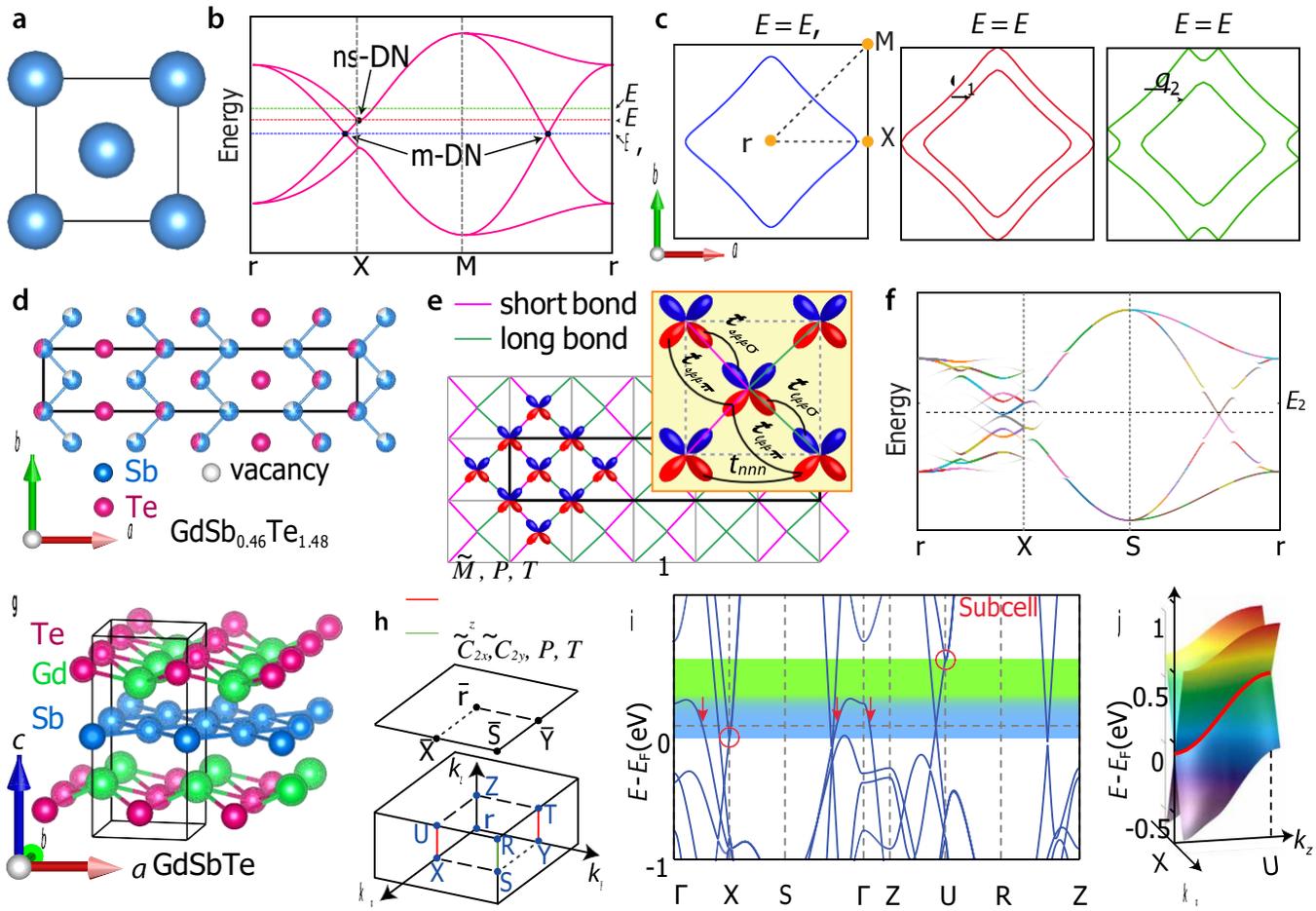

Figure 1: Electronic structure of GdSbTe and the effect of CDWs in combination with non-symmorphic symmetry. a) Illustration of a $4^4$-net lattice. b) Band structure for a four-band TB model. $E_1$, $E_2$, and $E_3$ represent Fermi levels with three different band-fillings; $E_1$ corresponds to half filling, $E_2$ cuts right through the four-fold band degeneracy at X, and $E_3 = 2E_2 - E_1$, the ns-DN and m-DNs are indicated. c) Fermi surface plots corresponding to the Fermi level at $E_1$, $E_2$, and $E_3$. $q_1$ and $q_2$ represent the CDW nesting vectors. d) Top view of the distorted square-net lattice forming a five-fold supercell in GdSb$_{0.46}$Te$_{1.48}$, with Te partial substitution of Sb, and the square net forming a chain-like texture. This illustrated supercell pattern is determined from single crystal diffraction [32]. e) Illustration of the TB model accounting for the five-fold superstructure, with $p_x$ and $p_y$-orbitals on each site. The short and long bonds are colored in purple and green, respectively. The top-right inset illustrates the definition of the 5 hopping parameters that are considered in the TB model. f) The calculated band structure from the superstructural TB model. $E_F$ is set so that it cuts through the ns-DN at X. Different bands from the supercell cell are illustrated in different colors. g) An illustration of the crystal structure of stoichiometric GdSbTe, which highlights the Sb $4^4$-net. h) An illustration of the BZ for space group *Pmmn*. The lines where four-fold degeneracy is enforced (in the presence of SOC) by a combination of a non-symmorphic and time-reversal symmetry are indicated. The top plane shows the (001) surface BZ. i) DFT-bulk band structure of stoichiometric GdSbTe without SOC. The energy span of the ns-Dirac line node along X-U is indicated by the colored window; the two endpoints at X and U are circled. The arrows indicate the trivial bands that cross the Fermi level, resulting in a hole-pocket at the FS for GdSbTe. For doped GdSb$_x$Te$_{2-x-\delta}$, this hole-pocket vanishes when the Fermi level lies within the green-shaded regime. j) Illustration of the ns-Dirac nodal line (colored in red) along X-U.

S3, Supporting Information). The constant energy contours (Figure 2a) reveal a *k*-dependent CDW gap, similar to that in other CDW-distorted square-lattice compounds, such as rare earth tritellurides *Ln*Te$_3$ [44, 45]. If we focus on the local *k*-region near U along UZ path, the effect can be seen down to 0.3 eV below the Fermi level, which results in a vague and incomplete diamond-shaped FS (Figure 2a). Without a CDW, a diamond shape would be the expected FS (Figure 1c). Once the band hybridization is turned on by the CDW, the FS will partially gap and therefore change shape. In related *Ln*Te$_3$ where the CDW transition temperature varies between 244 and 660 K [46, 47, 48], the maximum gap values were determined to be ∼0.4 eV [45]. The maximum gap in GdSb$_{0.46}$Te$_{1.48}$ should be comparable to, if not larger than, that in *Ln*Te$_3$, since the CDW in GdSb$_{0.46}$Te$_{1.48}$ melts at $T > 950$ K (Figure S4a, Supporting Information), which is significantly higher than those observed in *Ln*Te$_3$ phases. Based on the melting temperature, the *average* CDW gap can be estimated to be in the scale of ∼78 meV. At $E_F$, the FS (Figure 2c) only shows enhanced intensity at U and T, and very weak intensity close to U. This suggests that the majority of the diamond-shaped band crossings at the Fermi level are gapped, while states near the high symmetry points (U and T) where non-symmorphic degeneracies are enforced remain. Note that ARPES measurements are known to have a direct connection to the unfolded band structure, particularly for systems with enlarged cells with weak translational symmetry breaking, as is in the case of GdSb$_{0.46}$Te$_{1.48}$ [49, 50]. The spectral weight of the shadow bands, if any, reflects the strength of their coupling to the broken translational symmetry of the normal cell (in this case, the five-fold CDW distortion).

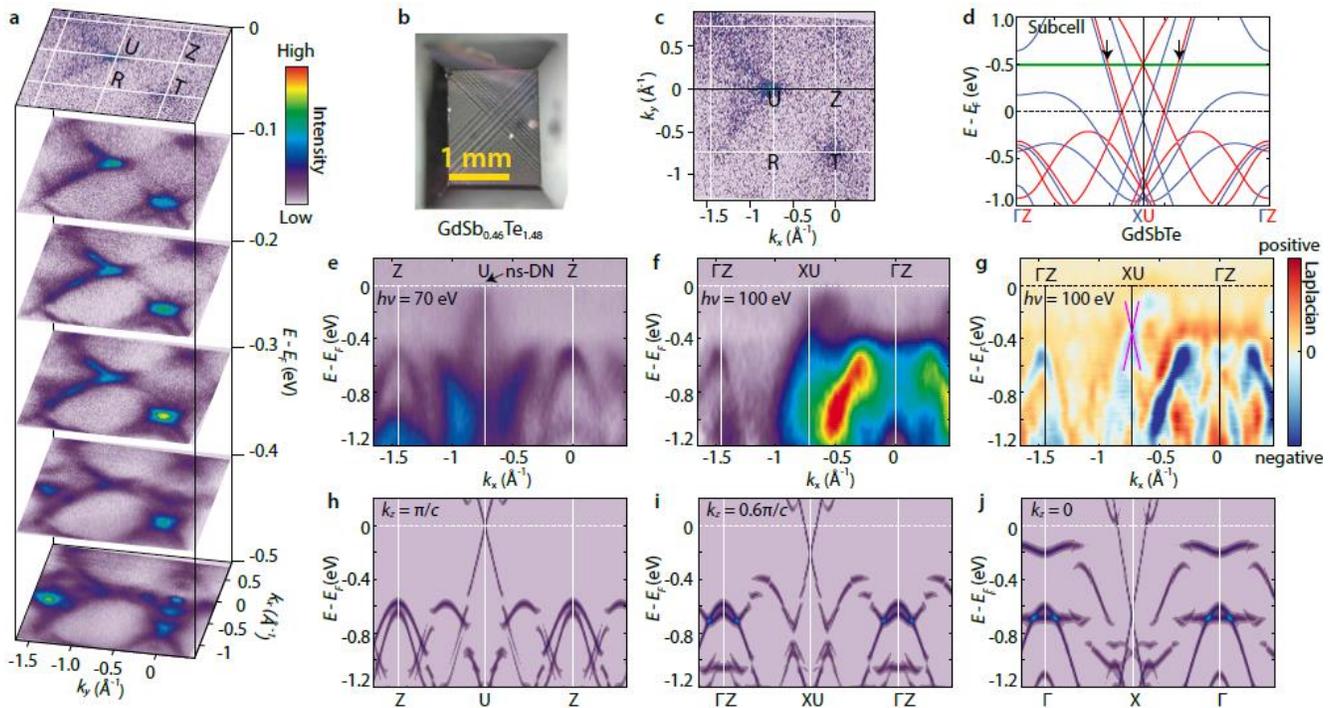

Figure 2: ARPES data taken at $hv$ = 70 and 100 eV, in comparison to DFT calculations. a) Constant energy plots at various initial state energies ranging from -0.5 to 0 eV, measured with a photon energy of $hv$ = 70 eV. b) A polarized optical image of a plate-like GdSb$_{0.46}$Te$_{1.48}$ crystal. c) Fermi surface measured at $hv$ = 70 eV. The black solid line indicates the line-path for band dispersion cut shown in (e). d) DFT band dispersion for the undistorted subcell along Γ-X-Γ overlapped with that along Z-U-Z. Arrows indicate the bands that are gapped by the CDW. Note that the green line is set to cut through the ns-DN at U, indicating the adjusted Fermi level of GdSb$_{0.46}$Te$_{1.48}$ for a direct comparison with the supercell DFT band structure in (h). e) Measured band dispersion along the Z-U-Z direction. The ns-DN at U is marked by the arrow. f) Measured band dispersion along the ΓZ-XU-ΓZ ($k_z$ = 0.6π/c plane) direction. g) Laplacian of the ARPES intensity plot shown in (f). Positive and negative Laplacians are a consequence of the minima and maxima, respectively. The Dirac crossing is marked by the purple lines, in comparison to the prediction in (i). h-j) DFT calculated band dispersion corresponding to $k_z$ = π/c, $k_z$ = 0.6 π/c and $k_z$ = 0 planes, respectively. The ARPES measurement at $hv$ = 70 eV was taken at 58 K, while the data at $hv$ = 100 eV was taken at 35 K. Both temperatures are above $T_N$ = 13.2 K.

To analyze the ARPES data, we performed DFT calculations on the superstructure, and unfolded the band structure to the subcell BZ [49] to allow a direct comparison with the measurements. Figure 2e and 2f show the measured band dispersion along Γ-X-Γ, which is parallel to the CDW direction. The momentum distribution curves and energy distribution curves are also plotted and shown in Figure S5, Supporting Information. The Z-U-Z cut (Figure 2e) measured with a photon energy of $h\nu$ = 70 eV shows a ns-Dirac crossing at U, with no other states interfering at the Fermi level. The overall band feature agrees with the DFT band dispersion in the $k_z = \pi/c$ plane (Figure 2h). Note the significant size of the band gap along this cut. We do not detect any clear feature of shadow bands. Figure 2f shows ARPES data measured at a photon energy of $h\nu$ = 100 eV, which corresponds to the band dispersion along ΓZ-XU-ΓZ at $k_z$ = 0.6 $\pi/c$ (Figure S3c, Supporting Information). A corresponding Laplacian plot [51], $\nabla^2 f = \frac{\partial^2 f}{\partial^2 x} + \frac{\partial^2 f}{\partial^2 y}$, where $f$ represents the measured ARPES intensity, and $x$ and $y$ represent the variables in $k$- and energy-space, respectively, is shown in Figure 2g. In agreement with DFT, the ns-DN lowers in energy moving from U to X along $k_z$. The measured ns-DN ($k_x = -\pi/a$, $k_y = 0$ and $k_z = 0.6\,\pi/c$) is at $E_i = -0.34$ eV, compared to $E_i = -0.21$ eV in the DFT calculation (Figure 2i). The calculated band dispersion along Γ-X-Γ ($k_z$ = 0 plane) is shown in Figure 2j. Here the ns-DN at X appears at $E_i = -0.66$ eV below the Fermi level. This indicates that the Dirac nodal line persists along the X-U line. If we compare the band structure of the supercell to that of the subcell (Γ-X-Γ in Figure 2d; for a more detailed comparison along all high-symmetry $k$-paths between subcell and supercell calculations, see Figure S6, Supporting Information), it becomes evident that bands contributing to the nodal line along X-U are not affected by the CDW, while the additional bands crossing the Fermi level along Γ-X are gapped.

We expect an in-plane anisotropy in the band structure due to the CDW, which is visible in the ARPES measurements. Figure 3a shows the ARPES band dispersion as well as its Laplacian (Figure 3b) along the Z-T-Z direction, perpendicular to the previous cut. The ns-DN at T is 0.29 eV below the Fermi level, in contrast to the near-Fermi-level ns-DN at U (Figure 2f). The experimentally measured band anisotropy is in qualitative, albeit not exact agreement with that from DFT, where the DN at T appears only 0.09 eV below the Fermi level (Figure 3c). We notice that twinning exists in the studied crystals, which can be seen in the polarized optical image shown in Figure 2b. However, twinning should not be responsible for this discrepancy. Orthorhombic twining could mix the signals from the Γ-X and Γ-Y directions, leading to a blurring of the ARPES band dispersion (another factor that may cause some blurring or band broadening is the band folding itself). Such blurring effects could lead to an underestimation of the in-plane anisotropy. However, the energy difference of 0.29 eV between the U- and T-DN is apparently much larger than the theoretical value of 0.09 eV. Therefore, we believe this difference may reflect the intrinsic discrepancy between experiment and DFT calculation.

Data for the diagonal cut along the RS-ZΓ-RS plane is also shown in Figure 3. Here, the CDW also gaps states at the Fermi level. The band dispersion plots measured with $h\nu$ = 100 eV are shown in Figure 3d (1st BZ) and 3g (2nd BZ) (data taken at $h\nu$ = 70 eV can be found in Figure S7 in the Supporting Information). Figure 3e and 3h show their Laplacian plots, respectively. At the Fermi level, the band intensity is strongly reduced. The DFT calculation (Figure 3f) reveals the cause for the vanishing intensity of the interfering bands: The CDW opens a gap along the SR-ΓZ-SR plane in BZ and the actual Fermi level is very close to the band edge (Figure 3f). At 0.62 eV below the Fermi level, the m-Dirac line nodes are visible in the DFT calculations. In the ARPES measurement, however, this crossing seems to be gapped by SOC, which appears at around 0.41 eV below $E_F$ (Figure 3h, indicated by black arrows). Comparing the two diagonal cuts in the 1st BZ and 2nd BZ, the latter is in good agreement with the DFT calculation, but for the former, a branch of the m-Dirac bands is missing (marked by dashed line). This phenomenon might be related to a matrix element effect [52]. The overall agreement between DFT and ARPES is high, albeit some disagreements in energy, which could be related to band renormalization due to correlation effects, which are not considered in the calculations, or other common inaccuracies of PBE-DFT. Such correlations could arise due to magnetic moments in the compounds or due to its nodal line structure, as correlations have been observed in the nonmagnetic square-net systems, ZrSiS [53] and ZrSiSe [54].

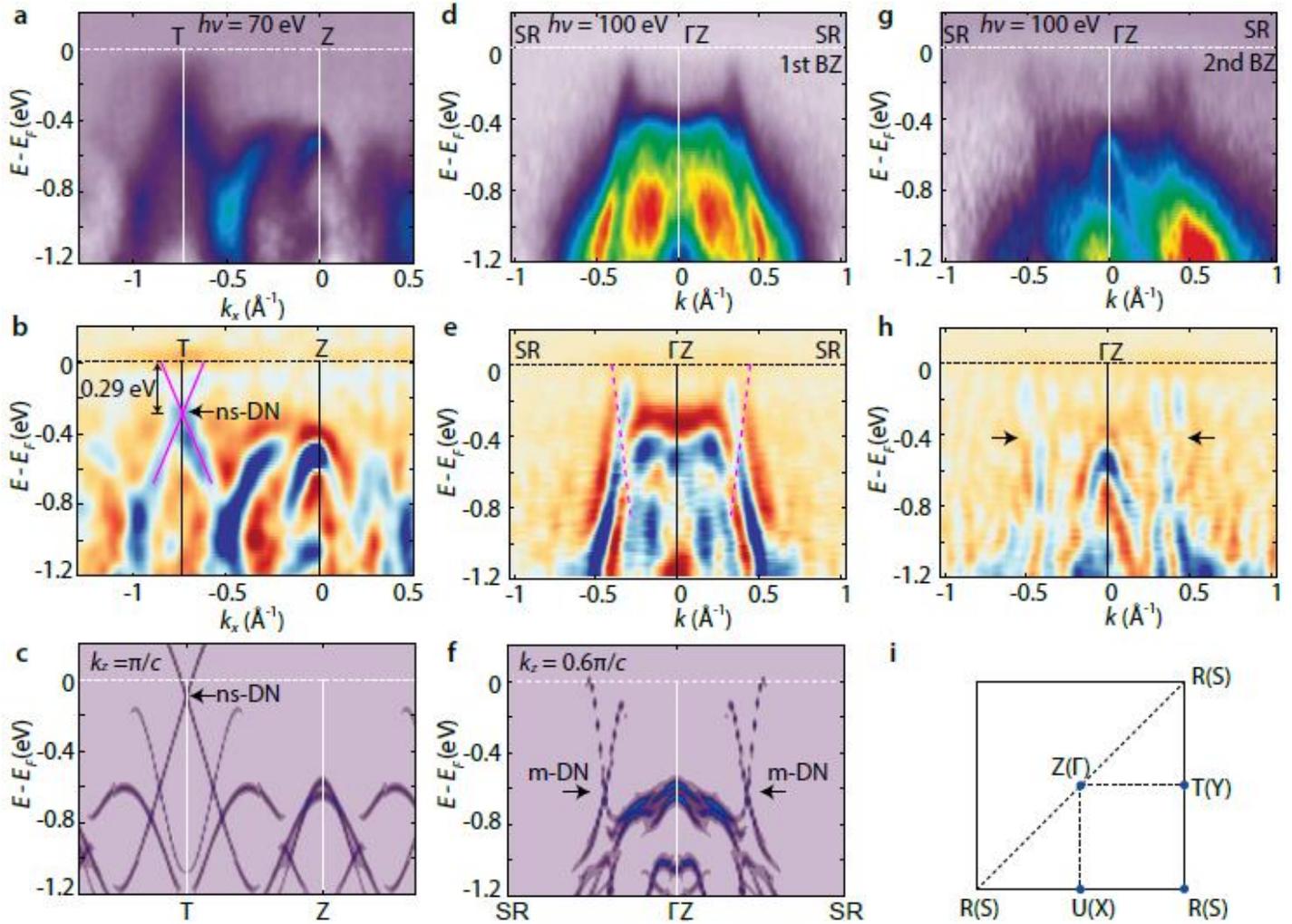

Figure 3: Band dispersion of GdSb$_{0.46}$Te$_{1.48}$ along further high-symmetry paths. a) ARPES intensity plot of the measured band dispersion along the Z-T-Z direction. b) Laplacian of the ARPES intensity plot shown in (a). The Dirac crossing is indicated by the purple lines, with the ns-DN at $E_i = -0.29$ eV. c) DFT calculated band dispersion at $k_z = \pi/c$. The DN is 0.09 eV below $E_F$. d,g) ARPES intensity plot of the measured band dispersion along the UR-ΓZ-UR direction ($k_z = 0.6\,\pi/c$ plane) in the 1st BZ and 2nd BZ, respectively. e,h) Laplacian of the ARPES intensity plots shown in (d) and (g), respectively. Arrows in (h) suggest the gapping of m-Dirac crossing due to SOC. The dashed purple lines in (e) indicate bands that are not clearly visualized in the 1st BZ, compared to that in 2nd BZ and the theoretical prediction in (f). f) DFT calculated band dispersion along SR-ΓZ-SR ($k_z = 0.6\,\pi/c$) (without SOC). Arrows indicate the m-DNs. i) Top view of the BZ in the planes of Z-U-R-T and Γ-X-S-Y.

Finally, we would like to point out that in DSMs that result from non-symmorphic symmetry such as ZrSiS, surface floating bands are expected to occur due to a reduced symmetry at the surface [55]. Thus the question whether such surface floating bands are observed in GdSb$_{0.46}$Te$_{1.48}$ naturally arises. As we show in Figure S8 (see Supporting Information), we observe surface states along Y – S – Y, which is the same direction floating bands are commonly observed as in other square net materials [55].
To assess the effect of a nearly isolated ns-DN at the Fermi level, we performed temperature dependent resistivity measurements on four compounds with varying Sb content: GdSb$_{0.42}$Te$_{1.46}$ (Fermi level slightly above the composition measured with ARPES), GdSb$_{0.46}$Te$_{1.48}$ (The same composition as the sample measured with ARPES), GdSb$_{0.57}$Te$_{1.40}$ (Fermi level lower than in the composition measured with ARPES), and GdSb$_{0.85}$Te$_{1.15}$ (tetragonal symmetry with no CDW or band folding). In our earlier discussion, we mentioned that there is a composition window (the corresponding Fermi energy window is indicated as a green shaded region in Figure 1i), where the Fermi level crosses the ns-Dirac nodal line along U-X. As our ARPES data indicates that the Fermi level crosses the ns-node at U in GdSb$_{0.46}$Te$_{1.48}$, we can infer that GdSb$_{0.57}$Te$_{1.40}$ should have Fermi level residing in the green region, while GdSb$_{0.42}$Te$_{1.46}$'s Fermi level is only slightly above the one measured with ARPES. Note that in this sample the Fermi level might be exactly at the ns-DN as the ARPES data could also be interpreted as having the Fermi level very slightly below the node. The temperature dependent resistivity curves of all samples are summarized in Figure 4a. In the case of tetragonal GdSb$_{0.85}$Te$_{1.15}$, the resistivity decreases as the temperature decreases until the magnetic ordering temperature $T_N$ = 12 K, which reflects clear metallic behavior. In contrast, in GdSb$_{0.42}$Te$_{1.46}$ and GdSb$_{0.46}$Te$_{1.48}$, the resistivity increases with a decrease in temperature, despite the non-zero density of states at the Fermi level. The resistivity behavior of GdSb$_{0.57}$Te$_{1.40}$ appears to show an intermediate behavior: as the sample cools down, the resistivity decreases until ~200 K, then increases until the lowest measured temperature. The temperature dependent resistivities of both GdSb$_{0.42}$Te$_{1.46}$ and GdSb$_{0.46}$Te$_{1.48}$ (in both samples the Fermi level is very close to the ns-DN at U) do not follow the classic activated behavior expected for semiconductors (Figure S9, Supporting Informa- tion). Above $T_N$, they can rather be described by power-law relations $\rho_1 T^{-0.45}$ and $\rho_2 T^{-0.29}$, respectively (Figure 4b). A power-law fit to the low-temperature resistivity of GdSb$_{0.57}$Te$_{1.40}$ is included in Figure 4b for comparison. As a reminder, in GdSb$_{0.57}$Te$_{1.40}$ the Fermi level is lower than in the samples measured with ARPES. The transport data (Figure 4b) indicates that GdSb$_{0.57}$Te$_{1.40}$ follows the power law in the smallest temperature range. Although the CDW also exists at room temperature in this compound, a clear deviation from the power law can be observed above ~90 K. Therefore, the existence of a CDW alone seems not to be the dominant reason for the power law behavior. This is different for the cases of GdSb$_{0.42}$Te$_{1.46}$ and GdSb$_{0.46}$Te$_{1.48}$. A good power-law fit can be obtained from ~20 K to room temperature. Hall measurements (Figure 4c) indicate that both GdSb$_{0.42}$Te$_{1.46}$ and GdSb$_{0.46}$Te$_{1.48}$ have very low carrier concentrations. The origin of the low carrier concentration should be attributed to the CDW-induced gap. We note that GdSb$_{0.42}$Te$_{1.46}$ has an incommensurate CDW wave vector (Figure S4b, Supporting Information), resulting in a more complicated FS than the commensurate GdSb$_{0.46}$Te$_{1.48}$. Nevertheless, the non-symmorphic symmetry protected Dirac nodes should persist, and it will be very close to the Fermi level. Based on this comparison study, we rule out that the CDW alone or the disorder effect cause the power-law resistivity behavior. Two ingredients seem to be important for this behavior: the existence of Dirac node very close to the Fermi level and minimal interfering bands.

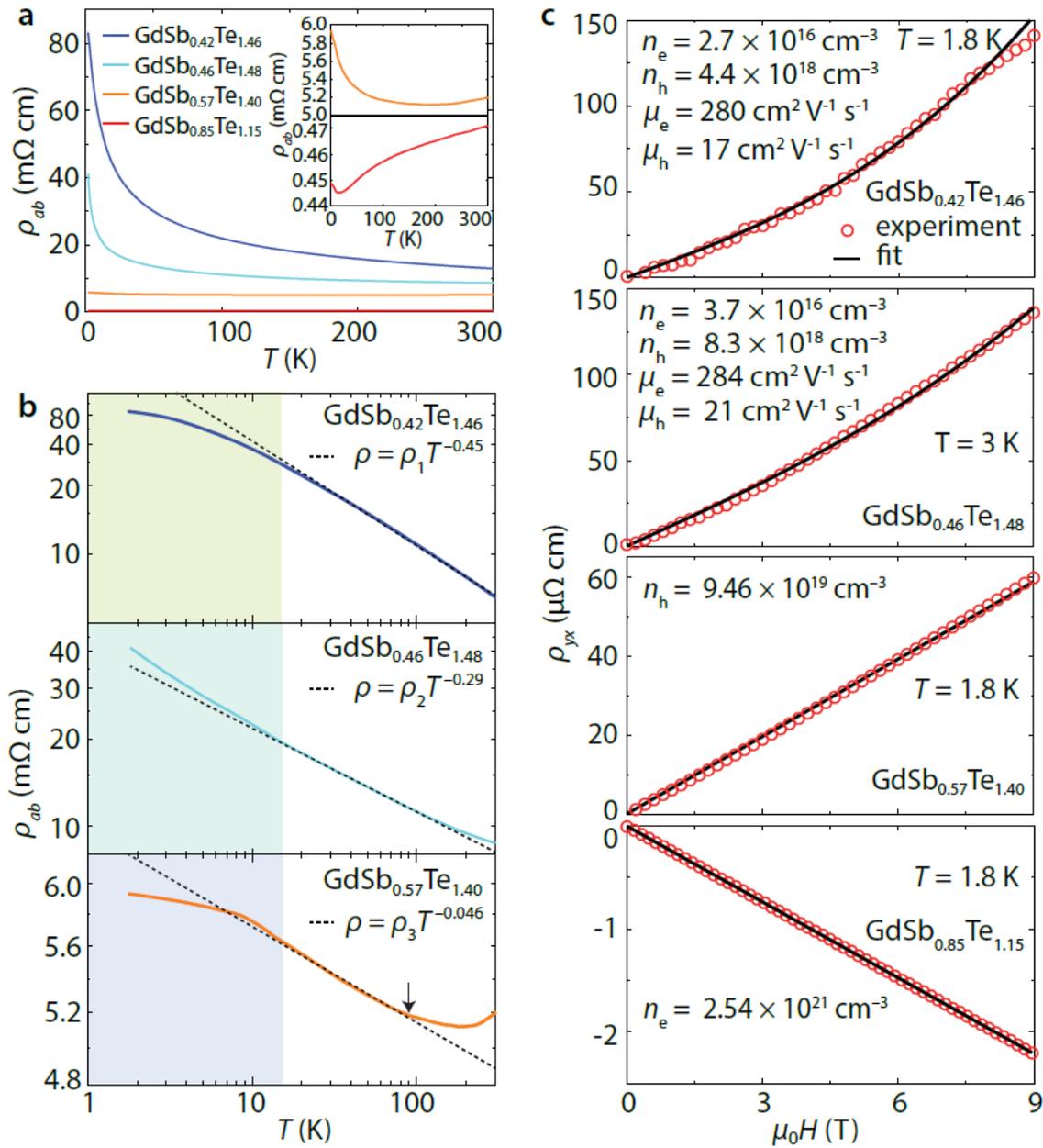

Figure 4: Transport properties of GdSb$_x$Te$_{2-x-\delta}$. a) Temperature dependent resistivity of GdSb$_x$Te$_{2-x-\delta}$. Inset shows the resistivities of only GdSb$_{0.57}$Te$_{1.40}$ and GdSb$_{0.85}$Te$_{1.15}$ for better visibility. b) The same data in a log-log plot. The dashed line shows a fit to the power-law in regime above the Néel temperature. The region of magnetic order is indicated by the shades. c) Low-temperature Hall resistivity of GdSb$_x$Te$_{2-x-\delta}$. For GdSb$_{0.42}$Te$_{1.46}$ and GdSb$_{0.46}$Te$_{1.48}$, two-band model fits are shown, while for GdSb$_{0.57}$Te$_{1.40}$ and GdSb$_{0.85}$Te$_{1.15}$, a one-band model was used.

The power-law resistivity behavior is perhaps the most interesting electronic property observed in GdSb$_x$Te$_{2-x-\delta}$. The gradual power-law increase reflects gradual localization of the carriers. However, such behavior seems incompatible with the standard picture of weak (Anderson) localization in either 2D or 3D. Anderson localization [56] occurs when the carrier's diffusion length $\mathcal{L} = \sqrt{D\tau_{in}}$ extends over a macroscopic distance of about 1 $\mu$m ($D$ is the diffusion constant and $\tau_{in}$ the inelastic lifetime determined by electron-phonon scattering). Within the coherent area $\mathcal{L}^2$ free from phonon scattering, constructive interference between two paths of a wave packet that are time-reversed partners leads to localization of the wave packet. Hence weak localization typically onsets at cryogenic temperatures or lower. By contrast, we observe a robust power law that extends to 300 K. At such warm temperatures, very strong phonon scattering renders $\tau_{in} < \tau$ ($\tau$ is the elastic scattering time), so that localization effects should not be observable. The magnetoresistance (MR) also strongly disagrees with weak localization. In Anderson localization, the resistivity increase observed at low $T$ is highly sensitive to suppression by a weak external magnetic field $H$ (the suppression is isotropic in 3D). The negative MR results from the destruction of the constructive interference within the coherent area. Complete suppression occurs when $H$ inserts a flux quantum $\varphi_0$ within the coherent patch. By contrast, the resistivity profile here is nearly insensitive to external $H$. In a 1 T field, $\rho$ increases, but only by 1.4% at 1.7 K (Figure S9b, Supporting Information). Therefore, although disorder in the material might play a role in transport properties, the observed power-law resistivity behavior does not agree with any known model with disorder playing the dominant role. A new transport model to assess the origin of this power-law behavior is thus required. We tentatively propose that the unusual situation of having Dirac nodes in the presence of CDWs, lattice disorder, and complex magnetic phases (as we show below) creates a new situation that has not yet been accessed experimentally. Further investigation of this regime is ongoing.

The Hall effect measurement shows reasonably good agreement with the ARPES measurements and DFT calculations. For GdSb$_{0.85}$Te$_{1.15}$, the Fermi level should lie within the blue-shaded regime in Figure 1i, where the corresponding FS consists of not only the nested diamond-shaped sheets, but also a hole-pocket centered at $\Gamma$ (Figure S2b, Supporting Information). Since the tetragonal GdSb$_{0.85}$Te$_{1.15}$ is determined to near the critical Sb/Te composition ratio beyond which a CDW distortion occurs (Figure S4b, Supporting Information), the Fermi level should be close to the blue-to-green transition level shown in Figure 1i. Therefore, the hole-carrier concentration described by the $\Gamma$-centered hole pocket in GdSb$_{0.85}$Te$_{1.15}$ would be negligible compared to the electron carriers described by the nested diamond-shaped sheets (Figure S2b, Supporting Information). At the critical point, the theoretical carrier concentration of electrons is calculated to be $1.7 \times 10^{21}$ cm$^{-3}$. This agrees reasonably well with the experimental observation that electrons are the dominant carriers as well as with the measured carrier concentration of $2.5 \times 10^{21}$ cm$^{-3}$ in GdSb$_{0.85}$Te$_{1.15}$. For the other three compounds, the Fermi surface is affected by the CDW-induced gap. In the ARPES measurement on GdSb$_{0.46}$Te$_{1.48}$, a vague incomplete diamond-shaped FS can only be revealed in the constant energy cut when the binding energy is ~0.1 eV. This picture is supported by the significant reduction of carrier concentration from Hall effect measurements in samples with a composition ranging from GdSb$_{0.85}$Te$_{1.15}$ to GdSb$_{0.46}$Te$_{1.48}$.

We also studied the low temperature magnetic phases of GdSb$_{0.46}$Te$_{1.48}$ with magnetic susceptibility measurements. Three magnetically ordered phases are visible (Figure S10, Supporting Information). Under a small field of 0.01 T, the highest transition appears at $T_N$ = 13.2 K, followed by a second transition at $T_1$ = 8.5 K, and third transition at $T_2$ = 7.2 K. When the magnetic field increases, the window between $T_1$ and $T_2$ shrinks and eventually disappears at a critical field of 1.1 T. Above this field, only two transitions exist up to 9 T. Overall, three magnetically ordered phases exist. Figure S10b, Supporting Information, shows an illustration of the magnetic phase diagram. Such a complex magnetic phase diagram is in sharp contrast to that of tetragonal GdSb$_{0.85}$Te$_{1.15}$, where only one magnetic transition was reported [32].

## 2 Conclusion

In summary, we introduced a new mechanism to design idealized non-symmorphic DSMs that is based on the cooperative effect of a CDW and non-symmorphic symmetry. The proposed mechanism is experimentally demonstrated in GdSb$_{0.46}$Te$_{1.48}$, showing that the band structure is composed of Dirac nodes at the Fermi level, with minimal interference of other states. Finally, we show that GdSb$_x$Te$_{2-x-\delta}$ possesses exotic transport behavior and complex magnetism. Arguably, the most interesting feature is the robust power-law increase in the resistivity which onsets near room temperature. From the high-temperature onset and the absence of significant MR, this "localization" effect does not follow the usual paradigm of Anderson localization. The unusual situation of having nearly isolated Dirac nodes in the presence of strong magnetic disorder could be the cause of this new transport regime deserving of further investigation.

The appearance of complex magnetism, CDWs, and Dirac states, as well as their overlap, in one material system is of great interest for future studies. This is in analogy to the interplay of structural, magnetic, and electronic degrees of freedom, that has long been the subject of study on high-temperature superconductors, such as La$_{1.6-x}$Nd$_{0.4}$Sr$_x$CuO$_4$ [57] and La$_{2-x}$Ba$_x$CuO$_4$ [58]. We note that the study of this interplay of magnetism, CDW, and Dirac states is not limited to the GdSb$_x$Te$_{2-x-\delta}$ system. The CDW instability is expected for electron-rich, layered square-net materials in general, that comply with the electron counting rules established by Papoian and Hoffmann [59]. Therefore, our proposed mechanism can be considered as a general mechanism to design an idealized non-symmorphic DSMs in this category of materials. In the LnSb$_x$Te$_{2-x-\delta}$ family, the essential requirement is the preservation of the non-symmorphic symmetry in the superstructure. We would like to note that incommensurate CDWs might also be able to preserve the Dirac nodes at the BZ boundary, while reducing additional band crossings at the Fermi level. Further studies are needed to elucidate this possibility. In this work, we did not yet analyze the role of magnetism on the band structure. Depending on the orientation of the spins, the electronic structure can be modified. Magnetism can be an additional way to tune the material properties, as has been reported for CeSbTe before [28].

## 3 Experimental Section

*Sample synthesis and characterization*: GdSb$_{0.46}$Te$_{1.48}$ single crystals were synthesized by chemical vapor transport, using iodine as the transport agent. For a detailed description of the synthesis procedure and composition characterization, see ref. [32]. The crystals are typically orthogonal structurally twinned, which could obscure some details in ARPES. ARPES and X-ray photoelectron spectroscopy (XPS) experiments were performed on in-situ cleaved crystals in ultrahigh vacuum (low $10^{-10}$ mbar). The ARPES spectra were recorded at 58 K with the $1^2$ ARPES experiment installed at the UE112-PGM2a beamline at the BESSY-II synchrotron, with various photon energies ($h\nu$) ranging from 58 eV to 100 eV. The ARPES measurement at $h\nu$ = 70 eV was taken at 58 K, while the ARPES at $h\nu$ = 100 eV was taken at 35 K. The core-level photoemission spectrum of GdSb$_{0.46}$Te$_{1.48}$ was measured at 14 K at the 29ID-IEX beam line (Advanced Photon Source, Argonne National Laboratory) using a hemispherical Scienta R4000 electron analyzer with a pass energy of 200 eV (energy and angular resolution are 220 meV and 0.1°, respectively). Resistivity and Hall measurement were performed on plate-like single crystal samples with patterned 6-terminal gold electrodes in a Quantum Design PPMS DynaCool system. A constant AC current with an amplitude of 5 mA was applied during the measurement. Temperature-dependent DC magnetization measurements were performed via the vibrating sample magnetometer (VSM) option in the same PPMS system. Temperature-dependent x-ray diffraction was performed at 11-BM at Argonne national lab on GdSb$_{0.46}$Te$_{1.48}$ powder from ground single crystals. Transmission electron microscopy (TEM) diffraction was performed on a double Cs-corrected Titan Cubed Themis 300 STEM equipped with an X-FEG source operated at 300 kV. The TEM samples were prepared by milling the bulk GdSb$_x$Te$_{2-x-\delta}$ single crystals using a focused ion beam.

*Electronic Structure Calculations*: A TB-model was constructed considering $p_x-$ and $p_y-$orbitals on each site, the same as that in refs. [27, 23]. Hoppings between nearest-neighbor (nn) and next-nearest-neighbor (nnn) orbitals were considered. For simplification, we consider only two types of nn-hopping. One for intra-chain hopping (short bonds): $t_{spp\sigma}$ = 1.5 eV and $t_{spp\pi}$ = $-$0.3 eV, and the other for inter-chain hoppings (long bonds): $t_{lpp\sigma}$ = 1.3 eV and $t_{lpp\pi}$ = $-$0.5 eV. The nnn-hopping is simplified with one parameter: $|t_{nnn}|$ = 0.11 eV. Note that this simplified treatment of the nnn-hopping does not change the symmetry of the system. The definitions of these parameters are illustrated in Figure 1e. The TB band unfolding is achieved by projecting the band eigenstates of a supercell Hamiltonian onto that of subcell [60, 61, 49] result, the intensity of the unfolded bands represent the spectral weight of each eigenstate with respect to the $p_x/p_y$ states in the non-distorted $4^4$-net.

Density functional theory (DFT) calculations were performed in VASP $v$5.4.4 [62, 63, 64] using the Perdew, Burke, Ernzerhof (PBE) functional [65]. PAW potentials [66, 67] were chosen based on the $v$ 5.2 recommendations. In order to study the role of the CDW, calculations were performed on a supercell with the experimentally-measured lattice parameters of GdSb$_{0.46}$Te$_{1.48}$, mimicking the true structure using the same five-fold lattice distortion along the $a$-axial direction. The Sb containing sites in GdSb$_{0.46}$Te$_{1.48}$ were modeled with full Sb occupancy, although the real structure contains partial vacancies and mixed Te occupancy as indicated in Figure 1d. For this reason, the DFT input used a hypothetical composition of GdSb$_{0.80}$Te$_{1.2}$, and the resulting Fermi level is lower than that of GdSb$_{0.46}$Te$_{1.48}$. The Fermi level was adjusted such that it crosses the ns-DN at X, to be consistent with our experimental observation and chemical intuition.

Self-consistent calculations for the DFT subcell were found to be well converged for a plane wave energy cutoff of 500 eV and a $k$-mesh density, $\pounds$ = 30 (corresponding to 7$\times$7$\times$3 and 1$\times$7$\times$3 Γ-centered $k$-meshes for the subcell and five-fold supercell, respectively); subsequent calculations were completed using settings equal to or better than these values. Localization of the Gd $f$ orbitals was corrected by applying a Hubbard potential $U$ = 6 eV using the method of Dudarev and coworkers [68]. Unfolded spectral functions for the supercell in the subcell BZ were calculated using the method of Popescu and Zunger [69] in VaspBandUnfolding [70]. Crystal structures were visualized with VESTA [71].

Note that both DFT subcell and supercell calculations were performed assuming a ferromagnetic (FM) order on the Gd lattice and neglecting valence spin-orbit coupling (SOC) effects. For a comparison with the experimental band structure in the paramagnetic state, we have ignored the small energy shift in the "up" and "down" spin channels, plotting only the down-spin bands. We find that SOC has almost no effect on the overall magnitude of the CDW-induced gap in the supercell calculation, although it introduces some small gaps at band-crossings and slightly shifts the relative energies of the majority and minority spin populations.

**Supporting Information**

Supporting Information is available from the Wiley Online Library or from the author.

**Acknowledgements**

This research was supported by the Arnold and Mabel Beckman Foundation through a Beckman Young Investigator grant awarded to L.M.S. The authors acknowledge the use of Princeton's Imaging and Analysis Center (IAC), which is partially supported by the Princeton Center for Complex Materials (PCCM), a National Science Foundation (NSF) Materials Research Science and Engineering Center (MRSEC; DMR-2011750). This research used resources of the Advanced Photon Source, a U.S. Department of Energy (DOE) Office of Science User Facility operated for the DOE Office of Science by Argonne National Laboratory under Contract No. DE-AC02-06CH11357. The work at UC Santa Barbara was supported by the National Science Foundation though the Q-AMASE-i Quantum Foundry, (DMR-1906325). We acknowledge use of the shared computing facilities of the Center for Scientific Computing at UC Santa Barbara, supported by NSF CNS-1725797, and the NSF MRSEC at UC Santa Barbara, NSF DMR-1720256. S.M.L.T. has been supported by the National Science Foundation Graduate Research Fellowship Program under Grant no. DGE-1650114. A.T. was supported by the DFG, proposal No. SCHO 1730/1-1. We thank HZB for the allocation of synchrotron radiation beamtime. M.K., D.M. and A.V. acknowledge support by the Impuls- und Vernetzungsfonds der Helmholtz-Gemeinschaft under grants No. HRJRG-408 and HRSF-0067 (Helmholtz-Russia Joint Research Groups). J.C. acknowledges the support from the National Science Foundation under Grant No. DMR-1942447 and the support of The Flatiron Institute, a division of the Simons Foundation. Any opinions, findings, and conclusions or recommendations expressed in this material are those of the authors and do not necessarily reflect the views of the National Science Foundation.